\documentstyle[preprint,revtex]{aps}
\begin{document}
\draft
\preprint{}
\begin{title}
Geometry, thermodynamics, and finite-size corrections \\
in the critical Potts model
\end{title}

\author{Chin-Kun Hu\cite{huck}, Jau-Ann Chen and N. Sh. Izmailian}
\begin{instit}
Institute of Physics, Academia Sinica, Nankang, Taipei 11529, Taiwan
\end{instit}
\centerline{and}
\author{P. Kleban}
\begin{instit}
Laboratory for Surface Science and Technology and Department of Physics and\\
Astronomy, University of Maine, Orono, ME 04469\\
% fax: 207-5812255
\end{instit}

\vskip 2 mm
\centerline{(Received 22 January 1999)}

\begin{abstract}
We establish an intriguing connection between geometry and
thermodynamics in the critical $q$-state Potts model on two-dimensional
lattices, using the $q$-state
bond-correlated percolation model (QBCPM) representation.
We find that the number of clusters $<N_c>$ of the QBCPM has an
energy-like singularity for $q \ne 1$, which is reached
and supported by exact results, numerical simulation, and scaling
arguments. We also establish that the finite-size correction to
the number of bonds, $<N_b>$, has no constant term and explains the
divergence of related quantities as $q \to 4$, the multicritical point.
Similar analyses are applicable to a variety of other systems.

\end{abstract}

\noindent{PACS numbers: 05.50+q, 64.60.Fr, 75.10-b}

\vskip 6 cm

\narrowtext
Percolation \cite{sa94} and the $q$-state Potts model (QPM) \cite{wu82}
are related to many interesting problems in mathematics and science
and are ideal models for studying critical phenomena
\cite{sa94,wu82,hu92,lpps92,fssf,ziff97,kz98,hcik}.
In recent years, much attention has been paid to universal quantities
at or near the percolation point, such as the
critical existence probability $E_p$ or crossing probability \cite{lpps92},
finite-size scaling functions \cite{fssf}, excess cluster numbers
\cite{ziff97,kz98}, etc.
In a recent Letter, Ziff {\it et al.} \cite{ziff97} calculated the
number of clusters per lattice site, $n$, in
percolation on two-dimensional lattices with $N$ lattice sites and
periodic boundary conditions (PBC).
They found that $n=n_c+b/N+\dots$, where $n_c$ is $n$ in the limit
$N \to \infty$ and $b$ is a positive universal constant that may be
calculated using conformal field theory (CFT) \cite{kz98}.
In this paper, we consider the $q$-state bond-correlated percolation model
(QBCPM) \cite{hu82t4} on planar lattices $G$ of $N$ sites
and $E$ bonds, which is equivalent to the
QPM on $G$; the numbers of bonds and clusters of a
subgraph $G'$ of $G$ are denoted by $N_b(G')$ and $N_c(G')$, respectively.
In the QBCPM, as in ordinary percolation, a natural focus on geometric
properties such as cluster number arises. However, the system also has
non-trivial thermodynamics, impelling an investigation of the connections
between geometry and thermal behavior.
In this work we concentrate on the critical Potts models for definiteness;
however, our methods are much more generally applicable, as pointed out below.

We address this question by
investigating the universal behavior of finite-size corrections (FSC).
We show by exact calculation that when $q \ne 1$, the FSC for $<N_c>$
is linearly related to the FSC for $<N_b>$, i.e. surprisingly,
the number of clusters has an energy-like singularity.
This is quite different from the case $q=1$, which is equivalent to
bond random percolation \cite{kf69} studied by Ziff et al. \cite{ziff97,kz98}.
Numerical simulation, scaling theory for the infinite system, and finite-size
scaling arguments verify and illuminate this conclusion.
The latter also implies that $<N_b>$ has no constant finite-size
scaling term at criticality, which we verify explicitly for the Ising model
on a square (sq) lattice.  We also find that the FSC of $<N_c+gN_b>$
and its higher cumulants (where $g$ is defined below, and $g = 1/2$ for a sq
lattice) diverge as $q \to 4$,
which is attributable to the onset of logarithmic corrections
at the multicritical point and is understandable from
a renormalization group (RG) picture.

Here we briefly review the connection between the QBCPM and the
QPM \cite{hu82t4,hu92rev}. In the QPM, each site of the lattice G is
occupied by a spin $s_i$ with spin components $-s,-s+1,...,s-1$, and
$s$, where $1 \le i \le N,  2s+1=q$,  and  $q$ is an integer. The Hamiltonian of the QPM
is given by
\begin{equation}
-H/k_B T =K\sum_{<i,j>}\delta(s_i,s_j)+B\sum_is_i,
\label{Ham}
\end{equation}
Here the first summation is a sum over all nearest neighbors,
$\delta(s_i,s_j)=1$ or 0 when $s_i=s_j$ or $s_i \ne s_j$, respectively,
$K=J/k_B T>0$ is the normalized NN coupling constant, and $B=h/k_B T$ is the
normalized external field with $k_B$ being the Boltzmann constant and $T$
being the absolute temperature.

Using the subgraph expansion of Eq. (\ref{Ham}), Hu has shown that phase
transitions of the QPM are percolation transitions of the QBCPM,
in which a subgraph $G'$ appears with the weight
\begin{equation}
\pi(G',p,q)=p^{N_b(G')}(1-p)^{E-N_b(G')}q^{N_c(G')},
\label{Weight}
\end{equation}
where $p=1-\exp{(-K)}=1$;
the spontaneous magnetization $M$ and the magnetic susceptibility $\chi$
of the QPM are related to the percolation probability $P$ and the mean
cluster size $S$ of the QBCPM, respectively. These
connections ensure that phase transitions of the QPM are percolation
transitions of the QBCPM \cite{hu82t4}.
The partition function of  the QPM at zero magnetic field may be
written as
\begin{equation}
Z_N=\sum_{G'} (\exp{(K)}-1)^{N_b(G')} q^{N_c(G')}=\exp{(KE)}\sum_{G'} \pi(G',p,q).
\label{ZN}
\end{equation}
Here the sum is over all $G'$ of $G$ \cite{kf69}.
The internal energy $U$ and the specific heat $C_h$ of the QPM
are related to the average number of occupied bond, $\bar p$, and
the fluctuations of the number of occupied bonds, $C_{2b}$,
of the QBCPM, respectively \cite{hu82t4}.

Using the Swendsen-Wang algorithm \cite{sw87}, we
calculate the average number of clusters per site $n$ of the critical
QBCPM on $L' \times L$ square lattices with PBC in both horizontal
and vertical directions; the number of spin components $q$ is an
input parameter taken to be 1, 2, 3, and 4. It should be noted
that $n$ in the limit $L',~L \to \infty$, denoted by $n_c$, follows from
exact results for the critical Potts free energy on several planar lattices
\cite{wu82,nc4}. We plot $n-n_c$ as a function of $1/L^2$ in Fig. 1 which
shows that the data for $q=1$ are on a linear curve.
The linear least-square fit of these data gives $n_c=0.09807(6)$
and the slope $b=0.884 \pm 0.002$, which
are consistent with the result of Ziff et al. \cite{ziff97,kz98}.
However, results for $q$=2, 3, and 4 are quite
different, namely the curves for $q \ge 2$ have negative slopes,
which suggests that the argument of Ziff {\it et al.} \cite{ziff97}
to relate  the slope $b$ to the average number of clusters wrapping
around the toroidal system is invalid and signals a new behavior
as we show below.

To understand the curves in Fig. 1 for $q \ge 2$, consider
the partition function $Z_c$ of the planar lattice QPM at the critical
point $p_c=1-e^{-K_c}$:
\begin{equation}
Z_c=\sum_{G'}[f(q)]^{N_b(G')} q^{N_c(G')}.
\label{Zc}
\end{equation}
Here $f(q)=e^{K_c}-1$ and is known exactly for square, planar triangular,
and honeycomb lattices \cite{wu82}; for sq lattice, $f(q)=\sqrt q$.
$Z_c$ is supposed to factor as $Z_c=Z_n Z_u$, where
$Z_n$ is a nonuniversal factor and the universal factor $Z_u$ gives FSC.
Exact results for $Z_u$ follow from the Coulomb gas formulas of
Di Francesco, Saleur and Zuber (DFSZ) \cite{fsz87}.
The cumulants $C_n$ of $N_c+gN_b$ are given by
$ C_n=[q (\partial / \partial q)]^n \ln Z_c$, where
$g=g(q)=qf'(q)/f(q)$ and is 1/2 for the sq lattice.
Since $Z_u=Z_u(L'/L)$, FSC's to $C_n$ are scale invariant. Thus,
as in \cite{kz98} for $q=1$
\begin{equation}
C_n= a_n LL'+b_n (L'/L)+O(1/L),
\label{Cn2}
\end{equation}
where $b_n$ is the universal FSC and
may be derived from DFSZ  \cite{fsz87}. It follows that there is no
divergent FSC term for any $C_n$ for $q < 4$. In particular,
for $n=1$ we have
\begin{equation}
C_1= <N_c+gN_b>=a_1 LL'+b(L'/L)+O(1/L).
\label{C1}
\end{equation}
For $q=2$, $b(1)=0.967734\dots$ and $b(2)=1.06463\dots$;
for $q=3$, $b(1)=1.05779\dots$ and $b(2)= 1.13321\dots$.
Since $<N_b>$ is proportional to the internal energy, which has a
singular FSC proportional to $L^{1/\nu}$ at criticality,
Eq.(\ref{C1}) implies that the FSC for $<N_c>$ has an
energy-like singularity with amplitude -$g$ times the amplitude
of $<N_b>$. A similar argument holds for any $C_n$, suggesting that
$N_c \approx -g N_b$ in the sense of FSC,
i.e. we can replace $N_c$ by $-g N_b$ to calculate any leading FSC.

This conclusion also follows from scaling for the infinite system.
The singular part of the free energy per site $f_s$ may be written as
$ f_s \approx A(q)[p-p_c(q)]^{2-\alpha (q)}$,
where $A(1)=0$ for (random) percolation and $\alpha$ is the specific
heat exponent.
Differentiating $f_s$ with respect to $q$, we find
$<N_c> \approx -A(q)p'_c(q)[p-p_c(q)]^{1-\alpha (q)}LL'$,
showing that $<N_c>$ is energy-like to leading order for $q \ne 1$.

The universal (singular) part of the free energy $F_u$ is defined above at
the critical point. According to finite-size scaling theory \cite{pf84},
$F_u$ also extends to large but finite systems near criticality, with
\begin{eqnarray}
F_u&=&L L' f_u \approx \psi ((\beta - \beta_c)L^{1/\nu})L L'
\nonumber \\
&\approx& B + (\beta-\beta_c)C L^{1/\nu}-
    {1 \over 2}(\beta-\beta_c)^2 D L^{2/\nu}+\dots.
\label{Fu}
\end{eqnarray}
Here $\beta_c$, $\nu$, $B$, $C$, and $D$ depend on $q$;
$B$, $C L^{1/\nu}$, and $D L^{2/\nu}$ determine the universal
(singular) terms in
the free energy,
internal energy, and specific heat at the critical point, respectively.
Let $x=e^{\beta J}-1$, then $x_c=f(q)$.
Using the total partition function $Z_N$ and
$<N_c>=q (\partial / \partial q)ln Z_N$,
$<N_b>=x (\partial / \partial x)ln Z_N$,
we find for $\beta = \beta _c$ that

\begin{equation}
<N_c(G')>
\approx n_c LL'-  {q f'(q) \over J (x_c+1)}C L^{1/\nu}+q B'(q),
\end{equation}

\begin{equation}
<N_b(G')>
\approx n_b LL' + {f(q) \over J (x_c+1)}C L^ {1/\nu}.
\label{Nb}
\end{equation}
Note that exact results for $n_b$ are available and $n_b=1$ for
the sq lattice Potts model for any $q$ \cite{wu82}. Therefore,
\begin{equation}
C_1= <N_c+g N_b>=a_1 LL'+ q B'(q),
\end{equation}
where $a_1=n_c+g n_b$, which agrees with Eq.(\ref{C1}) with
$b=q B'(q)$.
Note that Eq.(\ref{Nb}) also implies that there is no constant FSC to $<N_b>$.

It follows from finite-size scaling theory \cite{pf84} that
\begin{equation}
C_{2b} = <N_b^2>-<N_b>^2=n_{2b} LL'+c_2 L^{2/\nu}+\dots.
\label{Fb}
\end{equation}
The CFT result therefore suggests that
\begin{equation}
<N_c^2>-<N_c>^2=n_{2c} LL'+g^2 c_2  L^{2/\nu}+\dots,
\label{Fc}
\end{equation}

\begin{equation}
<N_cN_b>-<N_b><N_c>=n_{cb} LL'-g c_2  L^{2/\nu}+\dots.
\label{Fbc}
\end{equation}
It follows from Eqs. (\ref{Cn2}), (\ref{Fb}), (\ref{Fc}), and (\ref{Fbc})
that $ a_2=n_{2c}+g^2 n_{2b}+2g n_{cb}$.
Now we proceed to test the above predictions.

In \cite{ff69}, the internal energy of the Ising model on a large
$L' \times L$ square lattice at the critical point, $U_I(T_c)/J_I$, is given by
$ -U_I(T_c)/J_I =\sqrt 2 + 2 \Theta /L +d' /L^2+ \dots$, where
$ \Theta ={{\theta}_2{\theta}_3{\theta}_4}/({\theta}_2+{\theta}_3+
 {\theta}_4)$, $J_I$ is the coupling constant of Ising spins and is
related to $J$ by $J=2J_I$, $\theta_2$, $\theta_3$, and $\theta_4$ are
elliptic theta functions defined by Eq.(3.14) of \cite{ff69} and $d'$
was not determined in \cite{ff69}. We have extended the expansion
of $ -U_I(T_c)/J_I$ up to order $1/L^3$ and find

\begin{equation}
-\frac{U_I}{J_I} = \sqrt{2} + \frac{2}{L}\Theta
 -\frac{2}{L^3}\Theta \Theta_1 + O\left( \frac{1}{L^4}\right),
\label{Uee}
\end{equation}
where
\begin{equation}
\Theta =\frac{{\theta}_2{\theta}_3{\theta}_4}{{\theta}_2+{\theta}_3+
 {\theta}_4} \qquad {\mbox and} \qquad
\Theta_1=\frac{\pi^3 R}{96}\frac{{\theta}_2^9+{\theta}_3^9+{\theta}_4^9}{
{\theta}_2+{\theta}_3+{\theta}_4}
\label{theta1}
\end{equation}

The aspect ratio $(R)$ and the modulus $(k)$ of the complete elliptic
integrals of the first kind ($K(k)$)  are related to each other by
$R = K(k')/K(k)$ with $k'=\sqrt{1-k^2}$.

Equation (\ref{Uee}) shows that $d'$ is zero,
as predicted by Eq.(\ref{Nb}).  Since $U_I(T_c)=2J_I+U$ and
$U = - {\partial\over{\partial {\beta}}}~{\ln Z_N/N}
   = -{zJ \over 2p} ~\bar{p}$, $\bar p=<N_b>/E$ and $n=<N_c>/N$ are given by
\begin{equation}
\bar p = {1 \over 2}+ {p_c \Theta \over 2L}
  +{p_c \over 2L^3}\Theta \Theta_1
  +O\left( \frac{1}{L^3}\right), \label{pcE}
\end{equation}
\begin{equation}
n = n_c - {p_c \over 2}{\Theta \over L}
+ {b \over L^2} + O\left( \frac{1}{L^3}\right),
\label{ne}
\end{equation}
where $<N_b>$ and $<N_c>$ satisfy Eq.(\ref{C1}). As another test of
Eq.(\ref{Nb}), we plot our $\bar p$ data for square lattice three-state
Potts model as a function of $L^{1/\nu-2}=L^{-0.8}$ in Fig. 2,
which shows that the data fit a linear curve with slope
$s=0.1273\pm 0.0005$.
In Fig. 3, we plot $n-n_c$ data for the Ising model
and three-state Potts model as a function of $1/L$ for
$R=L'/L=$ 1 and 2. The solid lines represent  Eq. (\ref{ne}).
The dotted line represents $n-n_c=-s/L^{0.8}+b/L^2$
with $b=1.05779\dots$ obtained via \cite{fsz87}.
The agreement between numerical data and our predictions
is very good.

From \cite{ff69} and the connection between the specific heat and the
bond fluctuations, $C_{2b}$, of the QBCPM \cite{hu82t4}, we find that
at the critical point $p_c$ of the Ising model
$c_{2b}=C_{2b}/LL'=c_2 \ln{L} + n_{2b} +  O\left( \frac{1}{L^2}\right)$.
Here $c_2=2 p_c^2/\pi=0.218453\dots$,
$n_{2b}=\frac{1}{4}p_c^2 B(0,R)/K_I^2+1/(\sqrt{2}+1)$,
and $B(0,R)$ is defined by Eq.(4.21) of \cite{ff69}; for $R=1$,
$n_{2b}=0.475235\dots$.
Let $c_{2c}=(<N_c^2>-<N_c>^2)/LL'$ and
$c_{bc}=(<N_cN_b>-<N_b><N_c>)/LL'$.
For the Ising model (three-state Potts model), we fit $c_{2b}$, $c_{2c}$,
and $c_{bc}$ as linear functions of $\ln L$ ($L^{1/\nu-2}=L^{0.4}$)
to obtain $n_{2b}$, $n_{2c}$, and $n_{bc}$ and slopes.
$c_{2b}-n_{2b}$, $c_{2c}-n_{2c}$,
and $c_{bc}-n_{bc}$ for the $L \times L$ Ising model as a
function of $\ln L$ are shown in Fig. 4(a). The numerical values of
$c_2$ and  $n_{2b}$ are 0.21(8) and 0.47(6), respectively, which
are consistent with exact values.
The slopes for $c_{2c}$ and $c_{bc}$ are 0.06(0) and -0.11(5), respectively,
which are consistent with Eqs.(\ref{Fc}) and (\ref{Fbc}).
$c_{2b}-n_{2b}$, $c_{2c}-n_{2c}$, and $c_{bc}-n_{bc}$ for three-state
Potts model as a function of $L^{1/\nu-2}=L^{0.4}$
are shown in Fig. 4(b); the slopes of these curves are
0.64(3), 0.16(6), and -0.32(7), respectively, which are also consistent
with Eqs.(\ref{Fb})-(\ref{Fbc}).

As $q \to 4$, the system approaches a multicritical point.
From an RG point of view, its singular behavior may be understood
in terms of a dilution field $\psi$ and temperature field $\phi$ \cite{q4}.
Since $\psi \sim \epsilon=(4-q)^{1 \over 2}$, it follows from
scaling theory that $F_u$ will have an expansion in terms
with  integer powers of $\epsilon$ along the line of
critical points. Thus $b$, which is proportional to the $q$ derivative
of $F_u$, and all higher cumulants $b_n$ diverge as $q \to 4$.
This agrees with the results of a direct calculation using \cite{fsz87},
including the correct $\epsilon$ dependence. For the  cylinder geometry,
$b$ is finite but $b_n$ diverges for $n \ge 2$, which is attributable to
the vanishing of the leading term in the expansion of $F_u$ in
this geometry.

For $q=4$, one cannot derive results for the FSC to
$<N_c>$ by differentiation.
However, extending the scaling calculation in \cite{bn82}, we find
that to leading order

\begin{equation}
\bar p= <N_b>/E \approx 0.5+ A x (1-2 a \ln x)^{-3/4}=0.5+w(x),
\label{QQ4}
\end{equation}
where $x=L^{1/\nu-2}$ with $\nu=2/3$ for four-state Potts model \cite{wu82},
$A$ and $a$ are non-universal constants and the sq lattice bulk value
$\bar p = 0.5$ has been used.  The FSC part of this result
includes the effects of the constant term in the scaling relation for the
free energy \cite{bn82}. In Fig. 5, we plot data of $\bar p-0.5$ and $n-n_c$
for four-state Potts model \cite{nc4} as a function of $x=L^{-0.5}$.
Fitting $\bar p-0.5$ to $w(x)$ of Eq.(\ref{QQ4}) gives
$A=0.17\pm 0.01$ and $a=0.41\pm 0.05$. The solid and dotted lines
in Fig. 5 represent $w(x)$ and $-w(x)$, respectively.
Since $E = 2 LL'$ on the sq lattice, Fig. 5 shows that $-w(x)$ also
gives the leading FSC to $n-n_c$, which is
similar to the cases $q = 2$ and 3, i.e. we have numerical evidence for
the relation $N_c \approx -g N_b$ when $q =4 $.

Besides the Potts model, cluster representations are also useful for
understanding critical properties of a model of hydrogen bonding in water,
a dilute Potts model, the $O(n)$ model, quantum spin models, and many
others \cite{hu82t4,clusters}.  Our methods are useful for understanding
finite-size corrections in these systems.

We are indebted to I. Affeck, A. Aharony, J. L. Cardy, M. E. Fisher,
E. V. Ivashkevich, I. Peschel, P. Upton, F. Y. Wu and R. M. Ziff
for useful discussions.
This work was supported in part by the National Science Council of
the Republic of China (Taiwan) under grant
number NSC 88-2112-M-001-011.

\figure{$n-n_c$ as a function of $1 / L^2$ for the QBCPM
  on $L \times L$ sq lattices with pbc (torus) for $q=1$, 2,
  3, and 4.}

\figure{Numerical $\bar p$ of sq lattice three-state Potts model
as a function of $L^{1/\nu-2}$ with $\nu=5/6$ for three-state Potts model.
The solid line represents $\bar p =0.5 + s/L^{0.8}$ with
$s=0.127(3)$.}

\figure{Numerical $n-n_c$ of the $L' \times L$ sq lattice Ising model
and three-state Potts model as a function of $1 / L$.
The solid line represents Eq.(\ref{ne}) for the Ising model with
$b(1)=0.967734\dots$ and $b(2)=1.06463\dots$.
The dotted line represent the equation $n-n_c=-s/L^{0.8}+b/L^2$
for the three-state Potts model with $s=0.127(3)$ and $b=1.05779\dots$.}

\figure{(a) $c_{2b}-n_{2b}$, $c_{2c}-n_{2c}$, and
$c_{bc}-n_{bc}$ for the Ising model as a function of $ln L$,
 (b) $c_{2b}-n_{2b}$, $c_{2c}-n_{2c}$, and  $c_{bc}-n_{bc}$
 for the three-state Potts model as a function of $L^{2/\nu-2}=L^{0.4}$.}

\figure{$\bar p-0.5$ and $n-n_c$ for the four-state Potts model
  as a function of $x=L^{-1/2}$. The solid and dotted lines
 represents $w(x)$ and $-w(x)$, respectively.}


\begin{references}
\bibitem{nc4} For $q=4$, $n_c=0.16665\dots$ is obtained from
 the differentiation of exact critical free energy \cite{wu82} at $q=3.9999$.
\end{references}
\end{document}